\documentclass[11pt,a4paper]{article}
\usepackage{amsmath}
\usepackage{amssymb}
\usepackage{amsfonts}
\usepackage{amstext}
\usepackage{amsbsy}
\usepackage[mathscr]{eucal}
\usepackage{graphicx}
\usepackage{color}
\usepackage[all]{xy}
\newcommand{\beq}{\begin{equation}}
\newcommand{\eeq}{\end{equation}}
\newcommand{\beqa}{\begin{eqnarray}}
\newcommand{\eeqa}{\end{eqnarray}}

 \numberwithin{equation}{subsection}

\newcommand{\Tr}{\operatorname{Tr}}

\renewcommand{\phi}{\varphi}

\newcommand{\ket}[1]{|{#1}\rangle}

\newcommand{\obs}[1]{\hat{#1}}

\makeindex
\title{\Large\textbf{Geometry and structure of quantum phase space}}

\author{Hoshang Heydari\\\emph{ Department of Physics, Stockholm  University,} \\\emph{
SE-106 91 Stockholm, Sweden}}

\date{}

\begin{document}
\maketitle \thispagestyle{empty}\begin{abstract}
The application of geometry to physics has provided us with new insightful information about many physical theories such as classical mechanics, general relativity, and quantum geometry (quantum gravity). The geometry also plays an important role in foundations of quantum mechanics and quantum information. In this work we discuss a geometric framework for mixed quantum states represented by density matrices, where the quantum phase space of density matrices is equipped with a symplectic structure, an almost complex structure, and a compatible Riemannian metric. This compatible triple allow us to investigate arbitrary quantum systems. We will also discuss some applications of the geometric framework.
\end{abstract}
\section{Introduction}

Geometry effects real physical systems in all scales.  One could argue that the world has a well defined and rich geometrical structure. In general relativity the geometrical structures of physical systems are in one-to-one correspondence with the energy content of our universe. There is also hope that some day we are able to construct a quantum geometrical theory as a generalization of general relativity in order to unify it with quantum mechanics. However the standard formulation of quantum mechanics is algebraic. Thus a geometric formulation of general quantum mechanics is very important step toward a full description of our world as a quantum world. During recent decades we have witnessed emergent of a geometric formulation of quantum mechanics for pure states  \cite{Gunter_1977,Kibble_1979,Ashtekar_etal1998,Brody_etal1999}. However a geometric formulation of general quantum mechanics is still lacking. Recently we have also introduced a geometric framework for mixed quantum states based on principal fiber bundle that turn out very feasible and provides us with many results on foundations and applications of quantum mechanics \cite{GP,MB,DD,GQE,GUR,QSL}. We were also able to find yet another geometric formulation of quantum mechanics based on K\"{a}hler structure \cite{HH}.
In this contribution we will give a short introduction to geometric formulation of quantum mechanics. In particular, in section \ref{QPSPS} we review the geometry of quantum phase space of a pure state. In section \ref{QPSMS} we will give a formal definition of quantum phase space of mixed quantum states. Finally in section \ref{GSQPS} we will in details discuss the the geometrical structures of quantum phase space.

\section{Quantum phase space of pure states}\label{QPSPS}
A pure quantum state is defined on a complex projective space equipped with Fubini-Study metric such space usually called the quantum phase space of pure states.  The quantum phase space is a symplectic manifold which is equipped also with a symplectic structure and an almost complex structures. Thus the quantum phase space is a K\"{a}hler manifold. Let $\mathcal{H}$ be a Hilbert space. Then the set of states on $\mathcal{H}$ is defined by
\begin{equation}
\mathcal{S}(\mathcal{H})=\{\ket{\psi}\in\mathcal{H}:\langle\psi\ket{\psi}=1\}.
\end{equation}
However, the quantum states are defined on a projective Hilbert space $\mathcal{P}(\mathcal{H})=\mathcal{S}(\mathcal{H})/\sim$, where $\sim$ is an equivalent relation, that is quantum states are equivalent module a phase factor.  $\mathcal{P}(\mathcal{H})$ is called the quantum phase space and it is the space of one  dimensional projectors in $\mathcal{H}$. Note also that is also equivalent to the following $U(1)$ fiber bundle construction
 \begin{equation}
 U(1)\hookrightarrow\mathcal{S}(\mathcal{H})\longrightarrow\mathcal{P}(\mathcal{H}).
 \end{equation}
 If the Hilbert space is a finite dimensional Euclidean space $\mathcal{H}=\mathbb{C}^{n+1}$, then the quantum phase space $\mathcal{P}(\mathcal{H})=\mathbb{CP}^{n}$ is the complex projective space which is a K\"{a}hler manifold equipped with a canonical metric called Fubini-Study metric. In this case the fiber bundle is  the general Hopf fibration
   \begin{equation}
 U(1)\hookrightarrow \mathbb{S}^{2n+1}\longrightarrow\mathbb{CP}^{n}.
 \end{equation}
 The geometry of $\mathcal{P}(\mathcal{H})$ has been explored in many papers and books. But the geometry of mixed quantum states represented by density operator is still unexplored.  In the following section we will introduce the reader with a geometric framework that characterize mixed quantum states in simple effective way.
\section{Quantum phase space of mixed states}\label{QPSMS}
We will introduce a framework for density matrices based on idea of the co-adjoint orbit of a unitary matrix.
Let
\begin{equation}\label{spec}
\sigma=\left(
\nu_{1},m_{1};\nu_{2},m_{2};\ldots;\nu_{l},m_{l}
\right)
\end{equation}
be the spectrum of a density matrix  where $\nu_{1}<\nu_{2}<\ldots<\nu_{l}$ are the distinct eigenvalues of $\sigma_{i}$  with multiplicity $m_{i}$.  Then we define the space of density matrix by
\begin{eqnarray}
\nonumber
 D(\sigma)&=&\{\rho\in GL(\mathbb{C}):\rho=\rho^{\dagger},~~\text{with spectrum}~\sigma\}
\end{eqnarray}
which is an orbit of $U(n)$ action by conjugation and we assume that $\mathrm{Tr}(\rho)=1$. Thus the stabilizer of the elements of  $D(\sigma)$ are conjugate to
$U(m_{1})\times U(m_{2})\times\cdots\times U(m_{l})\subset U(n)$.
Thus the quantum phase space of density matrices $ D(\sigma)$ is a manifold that is diffeomorphic to the homogeneous space $U(n)/U(m_{1})\times U(m_{2})\times\cdots\times U(m_{l})$.

Next we argue that the quantum phase space  $ D(\sigma)$ is  actually a flag manifold.
 To see that let the sequence $K_{1},K_{2},\ldots,K_{l}$ be associated to the eigenspaces of $ D(\sigma)$, where $\dim K_{i}=m_{i}$ and $K_{i}\perp K_{j}$ for $i\neq j$. Now if we define $H_{i}=K_{1}\oplus K_{2}\oplus\cdots\oplus K_{i}$, then we have
 \begin{eqnarray}
\nonumber
0\subset H_{1}\subset H_{2}\subset\cdots\subset H_{l}=\mathbb{C}^{n}
\end{eqnarray}
which shows that $ D(\sigma)$ is a flag manifold. Next we give some examples to visualize the structures of the quantum phase space. The outermost orbits are given by the eigenvalues with multiplicity one. In this case the stabilizer is the equivalent to the torus $T^{n}=S^{1}\times S^{1}\times\cdots\times S^{1}$. We can also identify $K_{i}$ as a line and the quantum phase space become a complete flags, that is
 \begin{eqnarray}
\nonumber
D(\sigma)\cong \{0\subset H_{1}\subset H_{2}\subset\cdots\subset H_{l}=\mathbb{C}^{n}:\dim
H_{i}=i\}
\end{eqnarray}
with $\dim D(\sigma)=n^{2}-n$. The simplest non-trivial orbits are obtained when $l=2$ and the quantum phase space is identified with the grassmann manifold $G_{m_{i}}(\mathbb{C}^{n})$ of $m_{i}$-plane in $\mathbb{C}^{n}$ and $\dim D(\sigma)=2m_{1}m_{2}$. Thus $D(\sigma)$ is the homogeneous space $\frac{U(n)}{U(m_{1})\times U(m_{2})}$.  Note that the trivial case, namely when $\nu_{1}=1$ with multiplicity $m_{1}=1$ we get the quantum phase space of a pure state that we have discussed in previous section.
\section{Geometrical structures of quantum phase space}\label{GSQPS}
In this section we will introduce the geometrical structures of our quantum phase space including symplectic form, an almost complex structure and Riemannian  metric.
\subsection{Symplectic structures of quantum phase space}
First we equip the quantum phase space with a well defined symplectic structure. Let $\mathfrak{u}(n)$ be the Lie algebra of $U(n)$. Then the map $(X,Y)\mapsto \mathrm{Tr} (X^{\dagger}Y)$ defines a $\mathbb{R}$-linear form on $\mathfrak{u}(n)$ which is also invariant under conjugation and we can identify $\mathfrak{u}(n)$ with its dual vector space $\mathfrak{u}^{*}(n)$. Thus we can describe our quantum phase space as  coadjoint orbits of $\mathfrak{u}(n)$ which also posses a symplectic structure. Now, for any density matrix $\rho$ in $D(\sigma)$ we define a bilinear skew-symmetric form $\omega$ on the Lie algebra $\mathfrak{u}(n)$ as follows
\begin{equation}\label{KKS}
\omega(X,Y)=\mathrm{Tr}(\frac{1}{\imath \hbar}[X,Y]\rho).
\end{equation}
Suppose $\sigma\in \mathfrak{u}^{*}(n)$ and $D(\sigma)$ be the coadjoint orbit through $\sigma$. Then $D(\sigma)$ carries a symplectic structure $\omega(X,Y)$ defined by equation \ref{KKS}.

\subsection{Almost complex structure of quantum phase space}

We have shown that the quantum phase space $D(\sigma)$ is a flag manifold. In this section we will show that $D(\sigma)$ is a almost  complex manifold by defining an unique  almost complex structure on it. Let $\mathfrak{m}$ be a complex vector space. Then we write the Lie algebra as
 \begin{equation}
 \mathfrak{u}(n)=\mathfrak{g}\oplus \mathfrak{m}=\mathfrak{u}(m_{1})\oplus \mathfrak{u}(m_{2})\oplus\cdots\oplus \mathfrak{u}(m_{l})\oplus \mathfrak{m},
\end{equation}
where $\mathfrak{g}$ is a Lie subalgebra of $\mathfrak{u}(n)$ but $\mathfrak{m}$ in not. Thus we can decompose a matrix in an orbit as
\begin{equation}
\left(
  \begin{array}{ccccc}
    A_{1}1_{m_{1}} & X_{1,2} & X_{1,3} & \cdots & X_{1,l} \\
    -X^{\dagger}_{1,2} & A_{2}1_{m_{2}}  & X_{2,3} & \cdots & X_{2,l} \\
    \vdots &\cdots & \ddots & \cdots& \vdots \\
    -X^{\dagger}_{1,l-1}  & -X^{\dagger}_{2,l-1} & \cdots & \ddots & X_{l-1,l} \\
    -X^{\dagger}_{1,l}  & -X^{\dagger}_{2,l}  & \cdots & -X^{\dagger}_{l-1,l} & A_{l}1_{m_{l}}  \\
  \end{array}
\right),
\end{equation}
where $1_{m_{i}}$ are identity matrices of multiplicity $m_{i}$ and $A_{1}\in \mathfrak{u}(m_{i})$ and $X_{i,j}$ are complex matrices for all $i<j\leq l$.
The complex structure on $\mathfrak{m}$ is only depends on the complex structure of spaces of $X_{i,j} $. Thus we conclude that if
\begin{equation}
X=\left(
  \begin{array}{ccccc}
    0_{m_{1}} & X_{1,2} & X_{1,3} & \cdots & X_{1,l} \\
    -X^{\dagger}_{1,2} & 0_{m_{2}}  & X_{2,3} & \cdots & X_{2,l} \\
    \vdots &\cdots & \ddots & \cdots& \vdots \\
    -X^{\dagger}_{1,l-1}  &-X^{\dagger}_{2,l-1} & \cdots & \ddots & X_{l-1,l} \\
    -X^{\dagger}_{1,l}  & -X^{\dagger}_{2,l}  & \cdots & -X^{\dagger}_{l-1,l} & 0_{m_{l}}  \\
  \end{array}
\right)\in \mathfrak{m}
\end{equation}
where $0_{m_{i}}$ are zero matrices of multiplicity $m_{i}$, then the almost complex structure $J:\mathfrak{m} \longrightarrow \mathfrak{m}$ on $\mathfrak{m}$ is given by
\begin{equation}
j(X)=\left(
  \begin{array}{ccccc}
    0_{m_{1}} & \imath X_{1,2} & \imath X_{1,3} & \cdots &\imath  X_{1,l} \\
    - \imath X^{\dagger}_{1,2} & 0_{m_{2}}  & \imath X_{2,3} & \cdots & \imath X_{2,l} \\
    \vdots &\cdots & \ddots & \cdots& \vdots \\
    -\imath X^{\dagger}_{1,l-1}  &-\imath X^{\dagger}_{2,l-1} & \cdots & \ddots &\imath X_{l-1,l} \\
    -\imath X^{\dagger}_{1,l}  & -\imath X^{\dagger}_{2,l}  & \cdots & -\imath X^{\dagger}_{l-1,l} & 0_{m_{l}}  \\
  \end{array}
\right)\in \mathfrak{m}.
\end{equation}
The complex vector space $\mathfrak{m}$ can be considered as the tangent space to the homogeneous space $U(n)/U(m_{1})\times U(m_{2})\times\cdots\times U(m_{l})$ at the point image of $I\in U(n)$,  which is diffeomorphic to the quantum phase space, that is
\begin{equation}
\mathfrak{m}\simeq TD(\sigma)
\end{equation}
and it is equip with a complex structure. Now, let
$
\sigma$ as defined by equation (\ref{spec})
be a block diagonal matrix in the orbit of $D(\sigma)$ such that $\nu_{1}<\nu_{2}<\ldots<\nu_{l}$. Then any element in $TD(\sigma)$ can be written in a unique way as $[X,\sigma]$ for any $X\in \mathfrak{m}$.

\subsection{K\"ahler structure}

In this section we define  a specific K\"{a}hler form and derive an explicit expression for Hermitian inner product on the quantum phase space $D(\sigma)$. Moreover, we apply the geometric framework to derive a uncertainty relation for mixed quantum states.
The K\"{a}hler form on $D(\sigma)$ is defined by
\begin{equation}\label{KKSsymp}
\omega\left(\frac{1}{i\hbar}[\obs{A},\rho],\frac{1}{i\hbar}[\obs{B},\rho]\right)=\frac{1}{i\hbar}\Tr\left([\hat{A},\hat{B}]\rho\right).
\end{equation}
Note that
if $A$ is the expectation value function of a Hermitian operator $\obs{A}$, that is $A(\rho)=\Tr(\rho \obs{A})$, and $X_A$ is the Hamiltonian vector field associated with $A$, which is implicitly defined by the identity $dA(X)=\omega(X_A,X)$, then
\begin{equation}
X_A(\rho)=\frac{1}{i\hbar}[\hat{A},\rho].
\end{equation}
Now, $(\omega,J)$ is a K\"ahler structure, and we define $h$ to be the associated Hermitian inner product,
\begin{equation}
h(X,Y)=\omega(X,JY)+i\omega(X,Y)
\end{equation}
Next we will derive an expression for $h(X_A(\rho),X_B(\rho))$ where $\obs{A}$ and $\obs{B}$ are observables
which are off-diagonal at $\rho$:
\begin{equation*}
h(X_A(\rho),X_B(\rho))=\frac{2}{\hbar}\sum_{i>j} (\nu_i-\nu_j)\Tr(A_{ij}^\dagger B_{ij})
\end{equation*}
Our geometric framework  can be used to derive an uncertainty relation for mixed quantum states \cite{Hosh}.
Let $\hat{A}$  and  $\hat{B}$ be  observables on $\mathcal{H}$, and  consider the  uncertainty function
$\label{uf}
\Delta X(\rho)=\sqrt{\mathrm{Tr}(\rho\hat{S}^2)-\mathrm{Tr}(\rho\hat{S})^2},
$
where $\hat{S}$ can be either $\obs{A}$ or $\obs{B}$ . Then we have
\begin{equation}\label{gur}
\Delta A\Delta B\geq \frac{\hbar}{2}\sqrt{|h(X_A,X_B)|}.
\end{equation}
First we estimate $\Delta A(\rho)^2$ as follow
\begin{equation}
  \begin{split}
    \Delta A(\rho)^2
    &\geq \sum_{i>j}(\nu_i-\nu_j)\Tr(X_{ij}^\dagger X_{ij})
    =\frac{\hbar}{2}h(X_A(\rho),X_A(\rho)),
  \end{split}
\end{equation}
and similarly we get $\Delta B(\rho)^2\geq\frac{\hbar}{2}h(X_B(\rho),X_B(\rho))$.
Thus,
\begin{equation}
\begin{split}
\Delta A(\rho)^2\Delta B(\rho)^2
&\geq \frac{\hbar^2}{4}h(X_A(\rho),X_A(\rho))h(X_B(\rho),X_B(\rho))\\
&\geq \frac{\hbar^2}{4}|h(X_A(\rho),X_B(\rho))|,
\end{split}
\end{equation}
where in the last step we have used the Schwarz inequality.
Our geometric uncertainty relation are related to Robertson-Schr\"{o}dinger  uncertainty relation \cite{Robertson_1929}.

\section{Conclusion}
In this paper we have investigated the geometrical structure of quantum phase space of  mixed quantum states based on a K\"{a}hler structure. We have in details  discussed  a symplectic form, a Riemannian metric, and an almost complex structure on the quantum phase space.  Finally we have derived a geometric uncertainty relation for quantum assembles.  We also believe that our insightful result on quantum phase space of mixed quantum states can be applied to the  quantum mechanical systems with many applications in the fields of quantum dynamics, quantum information, and quantum computing.
\begin{flushleft}
\textbf{Acknowledgments:} The author acknowledges useful comments and discussions with Ole Andersson at Stockholm University and also discussion with Faisal Shah Khan at Khalifa University.
The  author also acknowledges the financial support from the Swedish Research Council (VR).
\end{flushleft}

\end{document}